\documentclass[printer]{aa}

\bibpunct{(}{)}{;}{a}{}{,}

\usepackage{graphicx}
\usepackage[varg]{txfonts}
\def\ks{km s$^{-1}$}
\def\d{$^\circ$}
\def\m{$^\prime$}
\def\s{$^{\prime\prime}$}
\def\hh{$^{\mathrm h}$}
\def\mm{$^{\mathrm m}$}
\def\ss{$^{\mathrm s}$}
\def\cm3{cm$^{-3}$}

\def\2{$^{12}$CO}
\def\3{$^{13}$CO}
\def\H{HCO$^{+}$}
\def\msol{M$_\odot$}

\begin{document}

\title{Discovering a misaligned CO outflow related to the red MSX source G034.5964-01.0292} 
\subtitle{}

\author {S. Paron \inst{1,2}
\and M. E. Ortega \inst{1}
\and A. Petriella \inst{1}
\and M. Rubio \inst{3}
}

\institute{Instituto de Astronom\'{\i}a y F\'{\i}sica del Espacio (IAFE),
             CC 67, Suc. 28, 1428 Buenos Aires, Argentina\\
             \email{sparon@iafe.uba.ar}
\and FADU and CBC - Universidad de Buenos Aires, Ciudad Universitaria, Buenos Aires
\and Departamento de Astronom\'{\i}a, Universidad de Chile, Casilla 36-D, Santiago, Chile
}

\offprints{S. Paron}

   \date{Received <date>; Accepted <date>}

\abstract{}{The red MSX source G034.5964-01.0292 (MSXG34), catalogued as a massive YSO, was observed in molecular lines with the aim of discover 
and study molecular outflows.}
{We mapped  a region of 3\m~$\times$ 3\m~centered at MSXG34 using the Atacama
Submillimeter Telescope Experiment in the \2 J=3--2 and HCO$^{+}$ J=4--3 lines with an angular and spectral resolution 
of 22\s~and 0.11 \ks, respectively. Additionally, public \3 J=1--0 and near-IR UKIDSS data obtained from the Galactic Ring Survey and the WFCAM 
Sciencie Archive, respectively, were analyzed.}
{We found that the \2 spectra towards the YSO present a self-absorption dip, as it is usual in star forming regions, and spectral wings evidencing 
outflow activity. The \H~was detected only towards the MSXG34 position at v$_{\rm LSR} \sim 14.2$ \ks, in coincidence with the \2 absorption dip and
approximately with the velocity of previous ammonia observations. \H~and NH$_{3}$ are known to be enhanced in molecular outflows.
Analyzing the spectral wings of the \2 line, we discovered misaligned red- and blue-shifted molecular outflows associated with MSXG34. 
The near-IR emission shows a cone-like shape nebulosity composed by two arc-like features related to the YSO, which can be due
to a cavity cleared in the circumstellar material by a precessing jet. This can explain the misalignment in the molecular outflows.
From the analysis of the \3 J=1--0 data we suggest that the YSO is very likely related to a molecular
clump ranging between 10 and 14 \ks. This suggests that MSXG34, with an associated central velocity of about 14 \ks, may be located in 
the background of this clump. Thus, the blue-shifted outflow is probably deflected by the interaction with dense gas along the line of sight.
From a SED analysis of MSXG34 we found that its central object should be an intermediate-mass protostar that probably will become a massive star.}{}

\keywords{Stars: formation -- ISM: jets and outflows -- ISM: molecules}

\maketitle

\section{Introduction}

One of the most outstanding and not completely understood process involved in the formation of stars is the 
appearance of collimated bipolar outflows in the earliest stages of formation. Such a process is present until the end of
the accretion phase producing significant consequences in the surroundings (e.g. \citealt{froe03a,froe03b,bally06,arce10}).
The structure and (a)symmetries of the outflows record orientation changes of the accretion disk and motion of the source
relative to the local interstellar medium \citep{cunni09}. Moreover, the outflows can be deflected by
material swept up in an earlier epoch of ejection by the central source \citep{fich97} and/or by dense preexisting molecular clumps \citep{choi05,baek09}.
Thus, mapping molecular outflows is very useful to study star formation, and in particular, to investigate the interaction between young stellar 
objects (YSOs) and the surrounding environments.

The red MSX source G034.5964-01.0292 (hereafter MSXG34), related to IRAS 1855+0056, is catalogued as a massive YSO located at the distance 
of 1.1 kpc \citep{lumsden13}. 
Ammonia was detected towards this source by \citet{wienen12}. The ({\it J},{\it K}) = (1,1) and (2,2) NH$_{3}$ 
inversion lines were detected at the velocities of 13.63 and 13.84 \ks, respectively, related to the ATLASGAL source G34.60-1.03 detected at 870 $\mu$m.
This source lies at the northeastern border of the HII region G34.5-1.1, that has a recombination line at 44.7 \ks~\citep{lockman89,kuchar97}, 
which discards any connection between them.
Thus, taking into account the presence of a likely massive YSO related to dense material traced by the ammonia emission, and that no 
others YSOs are catalogued around this source, which would make an outflow study confusing, we decided to observe MSXG34 in the \2 J=3--2 
and \H~J=4--3 lines to search for signatures of molecular outflows.

\section{Observations and data reduction}
\label{obs}

The molecular observations presented in this work were carried out on September 13, 2013
with the 10 m Atacama Submillimeter
Telescope Experiment (ASTE; \citealt{ezawa04}). We used the CATS345 receiver, which is a two-single
band SIS receiver remotely tunable in the LO frequency range of 324-372 GHz. We simultaneously
observed \2 J=3--2 at 345.796 GHz and HCO$^{+}$~J=4--3 at
356.734 GHz, mapping a region of 3\m~$\times$ 3\m~centered at RA $=$ 18\hh 58\mm 08.4\ss,
dec. $= +$01\d 00\m 38.8\s, J2000. The mapping grid spacing was 20\s~and the integration time was 20 sec in each pointing. 
All the observations were performed in position switching mode.

We used the XF digital spectrometer with bandwidth and spectral resolution set to
128 MHz and 125 kHz, respectively.
The velocity resolution was 0.11 \ks~and the half-power beamwidth (HPBW) 22\s.
The system temperature varied from T$_{\rm sys} = 150$ to 300 K. The main beam efficiency was $\eta_{\rm mb} \sim 0.65$.
The spectra were Hanning smoothed to improve the signal-to-noise ratio and only linear or some second order
polynomia were used for baseline fitting.
The data were reduced with NEWSTAR\footnote{Reduction software based on AIPS developed at NRAO,
extended to treat single dish data
with a graphical user interface (GUI).} and the spectra processed using the XSpec software
package\footnote{XSpec is a spectral line reduction package for astronomy which has been
developed by Per Bergman at Onsala Space Observatory.}.

Additionally, we used public \3 J=1--0 data, with an angular and spectral resolution of 46\s~and 0.2 \ks, respectively,
obtained from the Galactic Ring Survey (GRS; \citealt{jackson06}), and near-IR UKIDSS data \citep{lucas08} extracted from the WFCAM Science 
Archive\footnote{http://surveys.roe.ac.uk/wsa/}.

\section{Results and discussion}

\begin{figure}[h]
\centering
\includegraphics[width=6.5cm,angle=-90]{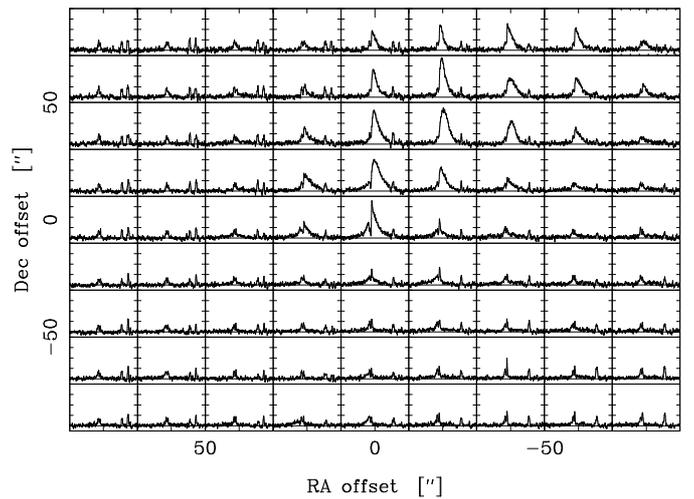}
\caption{\2 J=3--2 spectra obtained towards the surveyed region. The (0,0) offset is the position of studied source. }
\label{12cospectra}
\end{figure}

\begin{figure}[h]
\centering
\includegraphics[width=7cm]{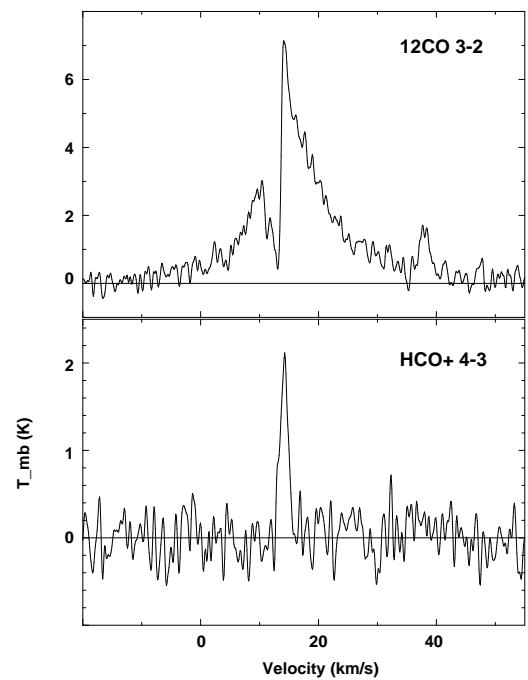}
\caption{\2 J=3--2 and \H~J=4--3 spectra obtained towards the center of the surveyed region. }
\label{spectra00}
\end{figure}

Figure \ref{12cospectra} displays the \2 J=3--2 spectra obtained towards the surveyed region. The center, i.e. the (0,0) offset, corresponds
to the position of MSXG34. The spectra present a main component with an absorption dip associated with MSXG34, 
and additionally less intense components at higher velocities, which appear in the whole surveyed area and represent gas detected along the 
line of sight not related to the analyzed source.
The same region was also surveyed in the \H~J=4--3 line, but emission was only detected at the (0,0) offset.
Figure \ref{spectra00} shows the \2 and \H~spectra towards the center of the region. 
The \2 J=3--2 spectrum shows a double peak structure like with a main component centered at $\sim$14.3 \ks~and a less intense 
component centered at $\sim$10.0 \ks. Both components are far to be gaussian, and it is very likely that the line appears 
self-absorbed as it is usually found towards star-forming regions (e.g. \citealt{johnstone03,buckle10}), which in this case, is evidenced by 
the absorption dip at $\sim$13.2 \ks. Moreover, the velocity of this absorption dip is almost coincident with those of the NH$_{3}$ lines detected
by \citet{wienen12} (v$_{\rm LSR} =$ 13.63 and 13.85 \ks~with $\Delta$v$_{\rm FWHM} =$ 0.87 
and 1.13 \ks~for ({\it J},{\it K}) = (1,1) and (2,2) NH$_{3}$
inversion lines, respectively). The presence of ammonia, tracer of high density gas, confirms the 
existence of a density gradient which produces the self absorption in the optically thick \2 line.
On the other side, the \H~spectrum shows a simpler behaviour and can be fitted with a gaussian centered at v$_{\rm LSR} \sim$14.2 \ks~with
$\Delta$v$_{\rm FWHM} \sim$1.6 \ks.

It is known that \H~and NH$_{3}$ enhance in molecular outflows \citep{torrelles92,girart98,raw04}. In effect, an enhancement in the abundance of such
molecular species is expected to occur in the boundary layer between the outflow and the surrounding
molecular core. As these authors point out, this enhancement would be due to the liberation and photoprocessing 
by the shock of the molecular material stored in the icy mantles of the dust.
This process may be occurring in MSXG34, which is supported by the observed complexity in the \2 J=3--2 profiles that, as shown in 
Figs. \ref{12cospectra} and \ref{spectra00}, present spectral wings as usually observed towards molecular outflows.

\begin{figure}[h]
\centering
\includegraphics[width=7.5cm]{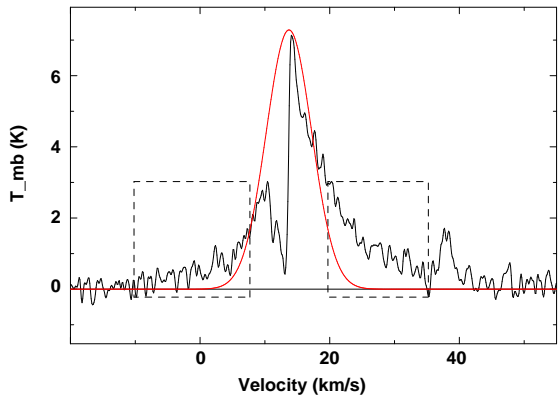}
\caption{\2 J=3--2 spectrum towards the center of the surveyed region with a gaussian function representing the main component not absorbed. 
The boxes show the high velocity gas intervals along which the emission was integrated. }
\label{gauss}
\end{figure}

\begin{figure}[h]
\centering
\includegraphics[width=8.5cm]{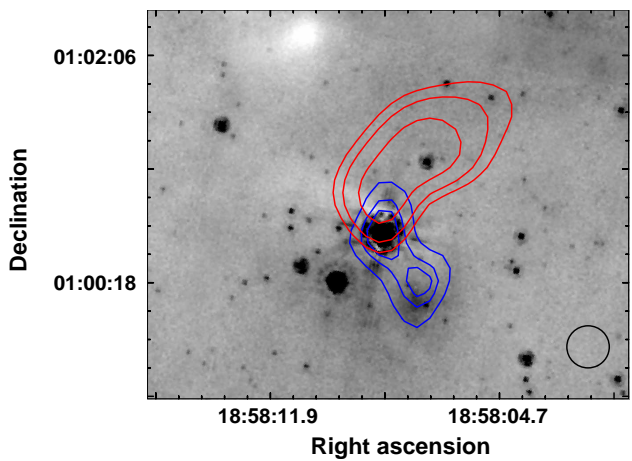}
\caption{{\it Spitzer}-IRAC 8 $\mu$m emission with contours of the \2 J=3--2 integrated between 20 and 35 \ks~(in red),
and between -10 and 7 \ks~(in blue). The contours levels are: 13, 17, and 25 K \ks~for the redshifted lobe, and 6.5, 7.5, and 8 K \ks~for 
the blueshifted one. The beam of the molecular observations is included in the bottom right corner.}
\label{lobes}
\end{figure}

Taking into account that \2 J=3--2 appears self-absorbed, a single gaussian function should contain the two 
main components that are separated by the absorption
dip. The emission that appears beyond the gaussian shape is considered to be associated with the molecular outflows.  
This is shown in Fig. \ref{gauss}, where 
over the central \2 spectrum are displayed the gaussian function and the high velocity gas intervals along which the emission was integrated. 
The small component at $\sim$ 38 \ks~is excluded because, as shown in Fig. \ref{12cospectra}, it appears almost in the whole surveyed region, 
which discards that it is produced by MSXG34. The result of the integration is shown in Fig. \ref{lobes}, 
where over the {\it Spitzer}-IRAC IR emission at 8 $\mu$m is displayed the \2 J=3--2 integrated between 20 and 35 \ks~(red-shifted gas), 
and between -10 and 7 \ks~(blue-shifted gas), respectively.
It can be seen a clear and intense red-shifted \2 lobe extending towards the northwest, while a less intense 
\2 blue-shifted lobe extends towards the southwest. Both lobes appears highly misaligned.

In order to roughly estimate the outflow mass, following \citet{bertsch93} we calculate the H$_{2}$ column density from
$$ {\rm N(H_{2})} = 2.0 \times 10^{20} ~\frac{\rm W(^{12}CO)}{\rm [K~km~s^{-1}]}~ {\rm (cm^{-2})}, $$
where W(\2) is the \2 J=3--2 integrated intensity along the velocity intervals shown in Fig. \ref{spectra00} (top). 
Then, the mass was derived from:
$$ {\rm M} = \mu~m_{{\rm H}} \sum_{i}{\left[ D^{2}~\Omega_{i}~{\rm N_{\it i}(H_{2}}) \right]}, $$
where $\Omega$ is the solid angle subtended by the beam size, $m_{\rm H}$ is the hydrogen mass,
$\mu$, the mean molecular weight, is assumed to be 2.8 by taking into account a relative helium abundance
of 25 \%, and $D$ is the distance. Summation was performed over all beam positions belonging to the lobes shown in Fig \ref{lobes}, 
giving the mass for the red- and blue-shifted outflows: M$_{red} \sim 7.5$ \msol~and M$_{blue} \sim 1.3$ \msol, respectively.
Then we obtain the momentum  P$_{red} \sim 114~\times~{\rm cos}^{-1}(\phi)$ M$_\odot$~km s$^{-1}$ and 
P$_{blue} \sim 22.5~\times~{\rm cos}^{-1}(\phi)$ M$_\odot$~km s$^{-1}$, and the energies
E$_{red} \sim 3.4 \times 10^{46}~ \times~{\rm cos}^{-2}(\phi)$ erg and E$_{blue} \sim 7.6 \times 10^{45}~ \times~{\rm cos}^{-2}(\phi)$ erg, 
where $\phi$ is the inclination angle of the outflow, which is uncertain.

\subsection{Outflows morphology}

As mentioned above and shown in Fig. \ref{lobes}, the red and blue molecular outflows are highly misaligned. 
A jet precession, produced by tidal interactions in a binary system or due to anisotropic accretion events (e.g. \citealt{papa95,kraus06}),
can generate misaligned molecular outflows. We found a similar case as presented here towards the UCHII region G045.47+0.05 \citep{ortega12}.
Two misaligned red- and blue-shifted molecular outflows were discovered through molecular observations obtained with ASTE. Later, from very
high-angular resolution observations at near-IR obtained with Gemini-NIRI, \citet{paron13} probed that a jet precession is occurring in a massive YSO.
Another similar case in which a jet precession scenario was suggested is in IRAS 20126+4104, where two misaligned CO high-velocity flows are observed
\citep{lebron06}. In order to investigate this possibility we analyze near-IR data from UKIDSS towards MSXG34.

\begin{figure}[h]
\centering
\includegraphics[width=9cm]{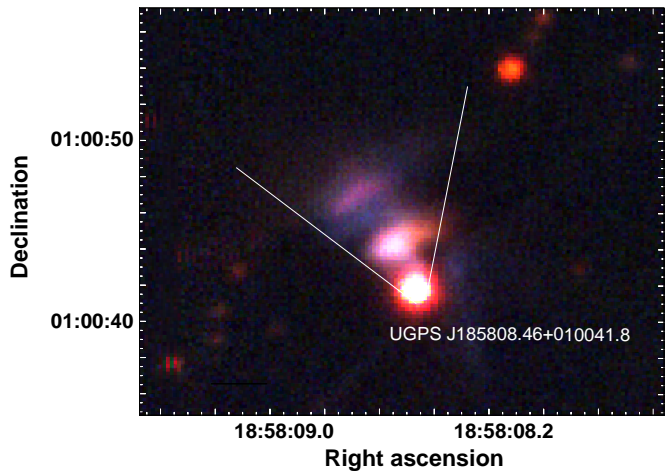}
\caption{Three-colour image towards MSXG34 where the {\it JHK}-bands from UKIDSS are presented in blue, green, and red, respectively. The
angular resolution is about 1 arcsec.}
\label{ukidss}
\end{figure}

Figure \ref{ukidss} shows the emission of the {\it JHK}-bands from UKIDSS in a three-colour image. 
It can be appreciated the point source UGPS J185808.46+010041.8 \citep{lucas08,lucas12}
that is connected with a cone-like shape nebulosity composed by two arc-like features, 
the closest one more intense than the farthest one. Both features seem to be connected by diffuse emission.
It is very likely that the near-IR emission forming a cone-like shape arises from a cavity cleared in the circumstellar
material. This emission can be due to a combination of different emitting processes: continuum
emission from the central protostar that is scattered at the inner walls of a cavity, emission from warm dust, and likely emission lines from 
shock-excited gas. The arc-like features, with a concavity pointing to the source, suggest an spiraling shape, which can be signature of a precessing jet.
This can be a similar case as we found in G045.47+0.05 \citep{paron13} and as others cases in the literature  (e.g. \citealt{wei06,kraus06}).
Therefore, we conclude that the misaligned CO outflows and the near-IR features related to the analyzed source strongly suggest a jet precession scenario.
However, higher-angular resolution observations in both, submillimeter and near-IR, are necessary to further investigate the whole circumstellar ambient.

\begin{figure}[h]
\centering
\includegraphics[width=9cm]{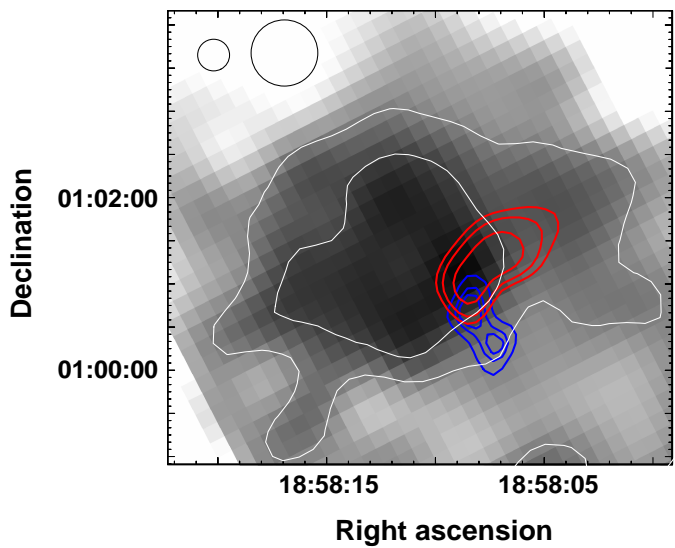}
\caption{\3 J=1--0 obtained from the GRS integrated between 10 and 14 \ks. The contours levels are 6.8 and 7.5 K \ks. The contours
of the blue and red lobes shown in Fig. \ref{lobes} are also displayed. The ASTE and GRS beams are included in the top left corner. }
\label{cloud13}
\end{figure}

Additionally, analyzing the molecular gas in a larger scale, we found that MSXG34 lies in a molecular clump. 
Thus, a complementary possible scenario for the misaligned molecular outflows is a deflection 
of one of its lobes produced by the interaction with dense material. By inspecting the \3 J=1--0 cube obtained from the GRS, 
we found that the molecular clump where MSXG34 is likely embedded extends from 10 to 14 \ks.
Taking into account that the systemic velocity of MSXG34 is about 14 \ks~(detemined from the emission of NH$_{3}$ and \H), 
the formation of this YSO would be occurring at the posterior border of the clump.
Figure \ref{cloud13} shows the \3 J=1--0 emission integrated between 10 and 14 \ks~with 
contours of the blue- and red-shifted molecular outflows. The scenario could be that the YSO is located in the background 
border of the molecular clump, the red-shifted lobe flows freely away, while the blue-shifted one hits the inner and densest portion 
of the clump deflecting its trajectory.
It could be a similar case as discovered in the NGC 1333 IRAS 4A region by \citet{choi05}. 
Submillimeter observations of high density tracers, like CS and SiO, 
are necessary to confirm the existence of high density gas belonging to the clump and to study the probable collision with the blue lobe.
Finally, another possibility is that the red- and blue-shifted lobes come from
monopolar molecular outflows, as proposed by \citet{fernandez13} in IRAS 18162-2048, where the respective counterlobes are not seen
because they are passing through a cavity and/or regions of low molecular abundance.

\subsection{On the nature of the outflow driving source}

\begin{figure}
\centering
\includegraphics[width=9cm]{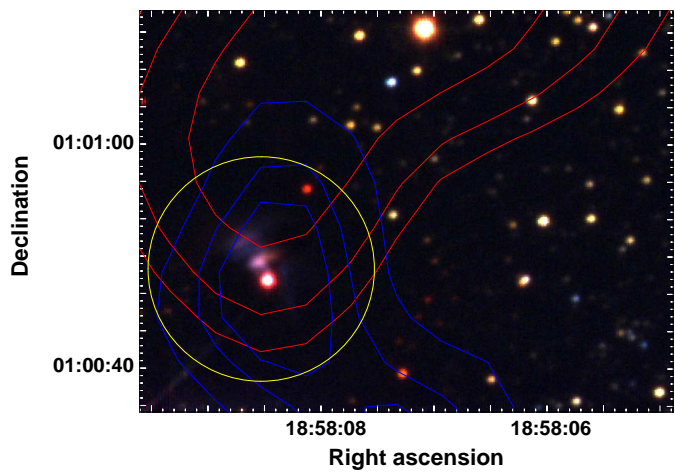}
\caption{UKIDSS three-colour image ({\it JHK}-bands in blue, green and
red, respectively) of the MSXG34 region. The red- and blue-shifted lobes of the molecular outflows are included.
The yellow circle, 30\s~in size, corresponds to the 500~$\mu$m continuum emission from SPIRE (above 5$\sigma$ of the rms noise level)
as extracted from the Herschel Science Archive.}
\label{ukidssconts}
\end{figure}

From the analysis of the UKIDSS data it is suggested that source UGPS
J185808.46+10041.8 is a good candidate to be the driving source of the
discovered molecular outflow (see Fig. \ref{ukidssconts}). This near-IR source is embedded in a clump
of cold dust catalogued as G034.60-1.030 in the ATLASGAL cold
high-mass clumps with NH$_3$ catalogue \citep{wienen12}. By considering
the associated dust continuum emission at 870~$\mu$m and following
\citet{beuther05} and \citet{hild83}, we estimate the mass of
this clump from:

\begin{eqnarray}
\tiny
M_{gas}=\frac{2.0 \times 10^{-2}}{J_{\nu}(T_{dust})}\frac{a}{0.1~
  \mu{\rm m}}\frac{\rho}{3~{\rm g~cm}^{-3}}\frac{R}{100}\frac{F_{\nu}}{{\rm
    Jy}}\nonumber\left(\frac{d}{{\rm kpc}}\right)^2
\left(\frac{\nu}{1.2~{\rm THz}}\right)^{-3-\beta}
\end{eqnarray}

\noindent where $J_{\nu}(T_{dust}$) = [exp($h\nu/kT_{dust})-1]^{-1}$
and $a, \rho, R,$ and $\beta$ are the grain size, grain mass density,
gas-to-dust ratio, and grain emissivity index for which we adopt the
values of 0.1~$\mu$m, 3~g~cm$^{-3}$, 100, and 2, respectively
(\citealt{hun97}, \citealt{hun00}, and \citealt{mol00}). Assuming a
dust temperature of 20~K and considering the integrated flux intensity
$F_{\nu} =$ 1.8~Jy at 870~$\mu$m \citep{wienen12}, we obtain $M_{gas} \sim 25$ \msol. 
By the other hand, considering the integrated flux density $F_{\nu} =$ 19.2~Jy at 500~$\mu$m obtained 
from the level 2 PLW Herschel image using the HIPE software package \citep{ott10}, and using the above equation with the same considerations, 
we estimate a mass for the clump of about 40 \msol. Thus, we conclude that UGPS J185808.46+10041.8 is embedded 
in a high mass clump (around 30 \msol).

\begin{figure}
\centering
\includegraphics[width=7.5cm]{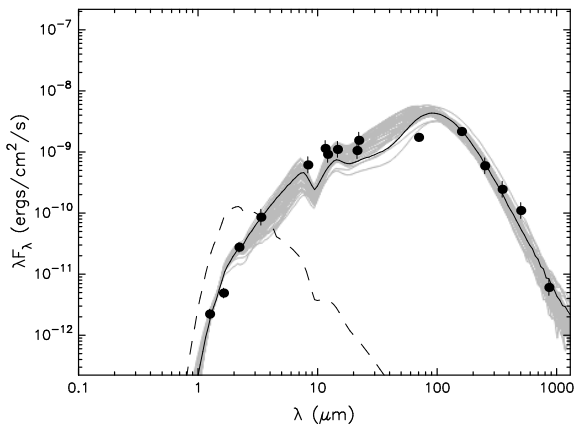}
\caption{SED of MSXG34. The circles indicate the measured fluxes of
the data points. The black and gray solid curves represent the
best-fitting model and the subsequent good-fitting models (with
$\chi^2 - \chi_{best}^2 \leq 3$), respectively. The dashed line
shows the stellar photosphere corresponding to the central source of
the best-fitting model, as it would look in the absence of
circumstellar dust.}
\label{sed}
\end{figure}

To better characterize the nature of this IR source, we performed a fitting 
of the spectral energy distribution (SED) using the online tool developed by
\citet{robi07}\footnote{http://caravan.astro.wisc.edu/protostars/}. We
adopt an interstellar extinction in the line of sight, A$_V$ , between
1 and 50 mag. We assume a 20~\% uncertainty for the distance to
UGPS J185808.46+10041.8. In Fig. \ref{sed} we show the SED with the 
best-fitting model (black curve), and the subsequent good-fitting models (gray curves)
with $\chi^2 - \chi_{best}^2 \leq 3$ (where $\chi_{best}^2$ is the
$\chi^2$ per data point of the best-fitting model for the source). To
construct this SED we use fluxes extracted from: UKIDSS-DR6 Galactic
Plane Survey \citep{lucas08} in the J, H and K bands (source UGPS
J185808.46+010041.8), MSX Point Source Catalog at 8.2, 12.1, 14.6, and 21.3 $\mu$m 
(source G034.5964-01.0292), WISE All-Sky Source Catalog\footnote{WISE is a
joint project of the University of California, Los Angeles, and the
Jet Propulsion Laboratory/California Institute of Technology, funded
by NASA.} at 3.6, 12, and 22 $\mu$m (source WISE
J185808.44+010041.8), PACS bands at 70 and 160 $\mu$m \citep{pogli10}, SPIRE bands at 250, 350, and 500~$\mu$m
\citep{griff10} from Herschel, and finally ATLASGAL at 870~$\mu$m.
PACS and SPIRE fluxes were obtained from level 2 MADmaps,  PLW, PMW, and PSW images, respectively, using
HIPE software package. Considering that UGPS
J185808.46+010041.8 is by far the brightest source within the Herschel
emission boundaries (see yellow circle in Fig. \ref{ukidssconts}) and no
contamination from clustering of infrared objects is observed in the
region, we treated the larger beam size data WISE at 22~$\mu$m,
Herschel, and ATLASGAL as ``data point'' instead of upper limits in
order to make the best use of all data in constraining the SED. 20\%
errors on the fluxes were assumed for all data, except for UKIDSS
fluxes where 30\% error were used due to extinction uncertainties.

The SED is fitted by multiple models, each model describing a set of
physical parameters. The same parameter from different models can have
a wide range spanning from several factors to orders of magnitudes.  We
obtained 63 good-fitting models that satisfy the $\chi^2$ criterion
above mentioned. In order to find a representative value for the
distributions of the parameters, we compute a weighted mean and a
range of values for some of the physical parameters of the source (see
Table \ref{sedtable}). The weight used for the weighted means is the
inverse of the $\chi^2$ of each model. It is important to mention that
the trend of our fitting results are not biased by the trend inherent
in the model grids. The SED analysis of UGPS J185808.46+010041.8
suggests that the central object is a young intermediate-mass
protostar of about 3 \msol. By comparing the obtained bolometric luminosity of 
about 200 L$_{\odot}$~with the total outflow mass, about 9 \msol, we note that it is 
in good agreement, within the dispersion, with the relation found by \citet{wu04}. The 
position of our point in the figure presented by the authors displaying M$_{\rm out}$ vs. 
L$_{\rm bol}$ also suggests an intermediate-mass protostar.

\begin{table}
\tiny
\caption{Main physical parameters from the SED of MSXG34.}\label{sedtable}
\begin{center}
\begin{tabular}{ccccccccc}
\hline
\hline
\multicolumn{2}{c}{$M_{\star}$}&\multicolumn{2}{c}{Age}&\multicolumn{2}{c}{$\dot
M_{env}$}&\multicolumn{2}{c}{$L$}\\
\multicolumn{2}{c}{[M$_{\odot}$]}&\multicolumn{2}{c}{[$\times
10^4$~yr]}&\multicolumn{2}{c}{ [$\times10^{-4}$ M$_{\odot}$yr$^{-1}$]}&\multicolumn{2}{c}{[$\times
10^{2} L_{\odot}$]} \\
\hline
Mean & Range & Mean & Range & Mean & Range & Mean & Range \\
3 & 1.5--8 & 1.0 & 0.1--3  & 6  & 1--10  & 1.9 & 0.5--5  \\
\hline
\end{tabular}
\end{center}
\end{table}

\section{Summary}

Using the ASTE telescope we observed MSXG34, a catalogued massive YSO at a distance of about 1 kpc, in the \2 J=3--2
and \H~J=4--3 lines with the aim of discover and study molecular outflows.
The \2 spectra towards the YSO present typical signatures of star forming regions: a self-absorption dip and spectral wings evidencing
outflow activity. The \H~was detected only towards the MSXG34 position at v$_{\rm LSR} \sim 14.2$ \ks, in coincidence with the \2 absorption dip and
approximately with the velocity of previous ammonia observations. The \H~and NH$_{3}$ are known to be enhanced in molecular outflows.

Analyzing the \2 J=3--2 emission we discovered misaligned red- and blue-shifted molecular outflows with mass, momentum and 
energy of: M$_{red} \sim 7.5$ \msol~and M$_{blue} \sim 1.3$ \msol,
P$_{red} \sim 114~\times~{\rm cos}^{-1}(\phi)$ M$_\odot$~km s$^{-1}$ and
P$_{blue} \sim 22.5~\times~{\rm cos}^{-1}(\phi)$ M$_\odot$~km s$^{-1}$, and 
E$_{red} \sim 3.4 \times 10^{46}~ \times~{\rm cos}^{-2}(\phi)$ erg and E$_{blue} \sim 7.6 \times 10^{45}~ \times~{\rm cos}^{-2}(\phi)$ erg,
where $\phi$ is the inclination angle of the outflow, which is uncertain.

Analyzing UKIDSS near-IR data we found 
that the emission shows a cone-like shape nebulosity composed by two arc-like features related to the YSO, which can be due
to a cavity cleared in the circumstellar material by a precessing jet, explaining in this way the misalignment in the molecular outflows.
Additionally, the \3 J=1--0 data show that MSXG34 is very likely embedded in a molecular clump that extends from 10 to 14 \ks.
Taking into account that the associated central velocity of MSXG34 is about 14 \ks, it is probable
that the YSO is located in the background of the clump densest portion, and thus the
blue-shifted outflow is probably deflected by the interaction with dense gas along the line of sight.
Another possibility is that the red- and blue-shifted lobes come from
monopolar molecular outflows, where the respective counterlobes are not seen
because they are passing through a cavity and/or regions of low molecular abundance.

Finally, we performed a SED analysis using fluxes from near- to far-IR, which suggests that the central object of MSXG34
is a young intermediate-mass protostar (about 3 \msol). The relation between the total outflow mass, obtained
from our molecular observations, and
the bolometric luminosity, obtained from the SED, also suggests an intermediate-mass stellar object.

\begin{acknowledgements}
We would like to thank the anonymous referee
for her/his helpful comments.
S.P. and M.O. are members of the {\sl Carrera del 
investigador cient\'\i fico} of CONICET, Argentina.
A.P. is a posdoctoral fellow of CONICET, Argentina. 
This work was partially supported by grants awarded by CONICET, ANPCYT and UBA (UBACyT) from Argentina.
M.R. wishes to acknowledge support from FONDECYT(CHILE) grant No1140839.
A.P. is very grateful to the ASTE staff for the support received during the observations.
The ASTE project is driven by Nobeyama Radio Observatory (NRO), a branch of
the National Astronomical Observatory of Japan (NAOJ), in collaboration with
University of Chile, and Japanese institutes including the University of Tokyo,
Nagoya University, Osaka Prefecture University, Ibaraki University, Hokkaido
University, and Joetsu University of Education.

\end{acknowledgements}

%%%%%%%%%%%%%%%%%%%%%%%%%%%%%%%%%%%%%%%%%%%%%%%%%%%%%%%%%%%%%%%%%%%%%
\bibliographystyle{aa} % style aa.bst
\bibliography{ref} % your references Yourfile.bib

\begin{thebibliography}{39}
\expandafter\ifx\csname natexlab\endcsname\relax\def\natexlab#1{#1}\fi

\bibitem[{{Arce} {et~al.}(2010){Arce}, {Borkin}, {Goodman}, {Pineda}, \&
  {Halle}}]{arce10}
{Arce}, H.~G., {Borkin}, M.~A., {Goodman}, A.~A., {Pineda}, J.~E., \& {Halle},
  M.~W. 2010, \apj, 715, 1170

\bibitem[{{Baek} {et~al.}(2009){Baek}, {Kim}, \& {Choi}}]{baek09}
{Baek}, C.~H., {Kim}, J., \& {Choi}, M. 2009, \apj, 690, 944

\bibitem[{{Bally} {et~al.}(2006){Bally}, {Licht}, {Smith}, \&
  {Walawender}}]{bally06}
{Bally}, J., {Licht}, D., {Smith}, N., \& {Walawender}, J. 2006, \aj, 131, 473

\bibitem[{{Bertsch} {et~al.}(1993){Bertsch}, {Dame}, {Fichtel}, {Hunter},
  {Sreekumar}, {Stacy}, \& {Thaddeus}}]{bertsch93}
{Bertsch}, D.~L., {Dame}, T.~M., {Fichtel}, C.~E., {et~al.} 1993, \apj, 416,
  587

\bibitem[{{Beuther} {et~al.}(2005){Beuther}, {Schilke}, {Menten}, {Motte},
  {Sridharan}, \& {Wyrowski}}]{beuther05}
{Beuther}, H., {Schilke}, P., {Menten}, K.~M., {et~al.} 2005, \apj, 633, 535

\bibitem[{{Buckle} {et~al.}(2010){Buckle}, {Curtis}, {Roberts}, \&
  {et~al.}}]{buckle10}
{Buckle}, J.~V., {Curtis}, E.~I., {Roberts}, J.~F., \& {et~al.} 2010, \mnras,
  401, 204

\bibitem[{{Choi}(2005)}]{choi05}
{Choi}, M. 2005, \apj, 630, 976

\bibitem[{{Cunningham} {et~al.}(2009){Cunningham}, {Moeckel}, \&
  {Bally}}]{cunni09}
{Cunningham}, N.~J., {Moeckel}, N., \& {Bally}, J. 2009, \apj, 692, 943

\bibitem[{{Ezawa} {et~al.}(2004){Ezawa}, {Kawabe}, {Kohno}, \&
  {Yamamoto}}]{ezawa04}
{Ezawa}, H., {Kawabe}, R., {Kohno}, K., \& {Yamamoto}, S. 2004, in Presented at
  the Society of Photo-Optical Instrumentation Engineers (SPIE) Conference,
  Vol. 5489, Society of Photo-Optical Instrumentation Engineers (SPIE)
  Conference Series, ed. J.~M. {Oschmann}, Jr., 763--772

\bibitem[{{Fern{\'a}ndez-L{\'o}pez} {et~al.}(2013){Fern{\'a}ndez-L{\'o}pez},
  {Girart}, {Curiel}, {Zapata}, {Fonfr{\'{\i}}a}, \& {Qiu}}]{fernandez13}
{Fern{\'a}ndez-L{\'o}pez}, M., {Girart}, J.~M., {Curiel}, S., {et~al.} 2013,
  \apj, 778, 72

\bibitem[{{Fich} \& {Lada}(1997)}]{fich97}
{Fich}, M. \& {Lada}, C.~J. 1997, \apjl, 484, L63

\bibitem[{{Froebrich} {et~al.}(2003{\natexlab{a}}){Froebrich}, {Smith}, \&
  {Eisl{\"o}ffel}}]{froe03a}
{Froebrich}, D., {Smith}, M.~D., \& {Eisl{\"o}ffel}, J. 2003{\natexlab{a}},
  \apss, 287, 217

\bibitem[{{Froebrich} {et~al.}(2003{\natexlab{b}}){Froebrich}, {Smith},
  {Hodapp}, \& {Eisl{\"o}ffel}}]{froe03b}
{Froebrich}, D., {Smith}, M.~D., {Hodapp}, K.-W., \& {Eisl{\"o}ffel}, J.
  2003{\natexlab{b}}, \mnras, 346, 163

\bibitem[{{Girart} {et~al.}(1998){Girart}, {Estalella}, \& {Ho}}]{girart98}
{Girart}, J., {Estalella}, R., \& {Ho}, P.~T.~P. 1998, \apjl, 495, L59

\bibitem[{{Griffin} {et~al.}(2010){Griffin}, {Abergel}, {Abreu}, {Ade},
  {Andr{\'e}}, {Augueres}, {Babbedge}, {Bae}, {Baillie}, {Baluteau}, {Barlow},
  {Bendo}, {Benielli}, {Bock}, {Bonhomme}, {Brisbin}, {Brockley-Blatt},
  {Caldwell}, {Cara}, {Castro-Rodriguez}, {Cerulli}, {Chanial}, {Chen},
  {Clark}, {Clements}, {Clerc}, {Coker}, {Communal}, {Conversi}, {Cox},
  {Crumb}, {Cunningham}, {Daly}, {Davis}, {de Antoni}, {Delderfield}, {Devin},
  {di Giorgio}, {Didschuns}, {Dohlen}, {Donati}, {Dowell}, {Dowell}, {Duband},
  {Dumaye}, {Emery}, {Ferlet}, {Ferrand}, {Fontignie}, {Fox}, {Franceschini},
  {Frerking}, {Fulton}, {Garcia}, {Gastaud}, {Gear}, {Glenn}, {Goizel},
  {Griffin}, {Grundy}, {Guest}, {Guillemet}, {Hargrave}, {Harwit}, {Hastings},
  {Hatziminaoglou}, {Herman}, {Hinde}, {Hristov}, {Huang}, {Imhof}, {Isaak},
  {Israelsson}, {Ivison}, {Jennings}, {Kiernan}, {King}, {Lange}, {Latter},
  {Laurent}, {Laurent}, {Leeks}, {Lellouch}, {Levenson}, {Li}, {Li},
  {Lilienthal}, {Lim}, {Liu}, {Lu}, {Madden}, {Mainetti}, {Marliani}, {McKay},
  {Mercier}, {Molinari}, {Morris}, {Moseley}, {Mulder}, {Mur}, {Naylor},
  {Nguyen}, {O'Halloran}, {Oliver}, {Olofsson}, {Olofsson}, {Orfei}, {Page},
  {Pain}, {Panuzzo}, {Papageorgiou}, {Parks}, {Parr-Burman}, {Pearce},
  {Pearson}, {P{\'e}rez-Fournon}, {Pinsard}, {Pisano}, {Podosek}, {Pohlen},
  {Polehampton}, {Pouliquen}, {Rigopoulou}, {Rizzo}, {Roseboom}, {Roussel},
  {Rowan-Robinson}, {Rownd}, {Saraceno}, {Sauvage}, {Savage}, {Savini},
  {Sawyer}, {Scharmberg}, {Schmitt}, {Schneider}, {Schulz}, {Schwartz},
  {Shafer}, {Shupe}, {Sibthorpe}, {Sidher}, {Smith}, {Smith}, {Smith},
  {Spencer}, {Stobie}, {Sudiwala}, {Sukhatme}, {Surace}, {Stevens}, {Swinyard},
  {Trichas}, {Tourette}, {Triou}, {Tseng}, {Tucker}, {Turner}, {Vaccari},
  {Valtchanov}, {Vigroux}, {Virique}, {Voellmer}, {Walker}, {Ward}, {Waskett},
  {Weilert}, {Wesson}, {White}, {Whitehouse}, {Wilson}, {Winter}, {Woodcraft},
  {Wright}, {Xu}, {Zavagno}, {Zemcov}, {Zhang}, \& {Zonca}}]{griff10}
{Griffin}, M.~J., {Abergel}, A., {Abreu}, A., {et~al.} 2010, \aap, 518, L3

\bibitem[{{Hildebrand}(1983)}]{hild83}
{Hildebrand}, R.~H. 1983, \qjras, 24, 267

\bibitem[{{Hunter}(1997)}]{hun97}
{Hunter}, T.~R. 1997, PhD thesis, Smithsonian Astrophysical Observatory, 60
  Garden St.~MS-78, Cambridge, MA 02178, USA

\bibitem[{{Hunter} {et~al.}(2000){Hunter}, {Churchwell}, {Watson}, {Cox},
  {Benford}, \& {Roelfsema}}]{hun00}
{Hunter}, T.~R., {Churchwell}, E., {Watson}, C., {et~al.} 2000, \aj, 119, 2711

\bibitem[{{Jackson} {et~al.}(2006){Jackson}, {Rathborne}, {Shah}, {Simon},
  {Bania}, {Clemens}, {Chambers}, {Johnson}, {Dormody}, {Lavoie}, \&
  {Heyer}}]{jackson06}
{Jackson}, J.~M., {Rathborne}, J.~M., {Shah}, R.~Y., {et~al.} 2006, \apjs, 163,
  145

\bibitem[{{Johnstone} {et~al.}(2003){Johnstone}, {Boonman}, \& {van
  Dishoeck}}]{johnstone03}
{Johnstone}, D., {Boonman}, A.~M.~S., \& {van Dishoeck}, E.~F. 2003, \aap, 412,
  157

\bibitem[{{Kraus} {et~al.}(2006){Kraus}, {Balega}, {Elitzur}, {Hofmann},
  {Preibisch}, {Rosen}, {Schertl}, {Weigelt}, \& {Young}}]{kraus06}
{Kraus}, S., {Balega}, Y., {Elitzur}, M., {et~al.} 2006, \aap, 455, 521

\bibitem[{{Kuchar} \& {Clark}(1997)}]{kuchar97}
{Kuchar}, T.~A. \& {Clark}, F.~O. 1997, \apj, 488, 224

\bibitem[{{Lebr{\'o}n} {et~al.}(2006){Lebr{\'o}n}, {Beuther}, {Schilke}, \&
  {Stanke}}]{lebron06}
{Lebr{\'o}n}, M., {Beuther}, H., {Schilke}, P., \& {Stanke}, T. 2006, \aap,
  448, 1037

\bibitem[{{Lockman}(1989)}]{lockman89}
{Lockman}, F.~J. 1989, \apjs, 71, 469

\bibitem[{{Lucas} {et~al.}(2008){Lucas}, {Hoare}, {Longmore}, {Schr{\"o}der},
  {Davis}, {Adamson}, {Bandyopadhyay}, {de Grijs}, {Smith}, {Gosling},
  {Mitchison}, {G{\'a}sp{\'a}r}, {Coe}, {Tamura}, {Parker}, {Irwin}, {Hambly},
  {Bryant}, {Collins}, {Cross}, {Evans}, {Gonzalez-Solares}, {Hodgkin},
  {Lewis}, {Read}, {Riello}, {Sutorius}, {Lawrence}, {Drew}, {Dye}, \&
  {Thompson}}]{lucas08}
{Lucas}, P.~W., {Hoare}, M.~G., {Longmore}, A., {et~al.} 2008, \mnras, 391, 136

\bibitem[{{Lumsden} {et~al.}(2013){Lumsden}, {Hoare}, {Urquhart}, {Oudmaijer},
  {Davies}, {Mottram}, {Cooper}, \& {Moore}}]{lumsden13}
{Lumsden}, S.~L., {Hoare}, M.~G., {Urquhart}, J.~S., {et~al.} 2013, \apjs, 208,
  11

\bibitem[{{Molinari} {et~al.}(2000){Molinari}, {Brand}, {Cesaroni}, \&
  {Palla}}]{mol00}
{Molinari}, S., {Brand}, J., {Cesaroni}, R., \& {Palla}, F. 2000, \aap, 355,
  617

\bibitem[{{Ortega} {et~al.}(2012){Ortega}, {Paron}, {Cichowolski}, {Rubio}, \&
  {Dubner}}]{ortega12}
{Ortega}, M.~E., {Paron}, S., {Cichowolski}, S., {Rubio}, M., \& {Dubner}, G.
  2012, \aap, 546, A96

\bibitem[{{Ott}(2010)}]{ott10}
{Ott}, S. 2010, in Astronomical Society of the Pacific Conference Series, Vol.
  434, Astronomical Data Analysis Software and Systems XIX, ed. Y.~{Mizumoto},
  K.-I. {Morita}, \& M.~{Ohishi}, 139

\bibitem[{{Papaloizou} \& {Terquem}(1995)}]{papa95}
{Papaloizou}, J.~C.~B. \& {Terquem}, C. 1995, \mnras, 274, 987

\bibitem[{{Paron} {et~al.}(2013){Paron}, {Fari{\~n}a}, \& {Ortega}}]{paron13}
{Paron}, S., {Fari{\~n}a}, C., \& {Ortega}, M.~E. 2013, \aap, 559, L2

\bibitem[{{Poglitsch} {et~al.}(2010){Poglitsch}, {Waelkens}, {Geis},
  {Feuchtgruber}, {Vandenbussche}, {Rodriguez}, {Krause}, {Renotte}, {van
  Hoof}, {Saraceno}, {Cepa}, {Kerschbaum}, {Agn{\`e}se}, {Ali}, {Altieri},
  {Andreani}, {Augueres}, {Balog}, {Barl}, {Bauer}, {Belbachir}, {Benedettini},
  {Billot}, {Boulade}, {Bischof}, {Blommaert}, {Callut}, {Cara}, {Cerulli},
  {Cesarsky}, {Contursi}, {Creten}, {De Meester}, {Doublier}, {Doumayrou},
  {Duband}, {Exter}, {Genzel}, {Gillis}, {Gr{\"o}zinger}, {Henning},
  {Herreros}, {Huygen}, {Inguscio}, {Jakob}, {Jamar}, {Jean}, {de Jong},
  {Katterloher}, {Kiss}, {Klaas}, {Lemke}, {Lutz}, {Madden}, {Marquet},
  {Martignac}, {Mazy}, {Merken}, {Montfort}, {Morbidelli}, {M{\"u}ller},
  {Nielbock}, {Okumura}, {Orfei}, {Ottensamer}, {Pezzuto}, {Popesso},
  {Putzeys}, {Regibo}, {Reveret}, {Royer}, {Sauvage}, {Schreiber}, {Stegmaier},
  {Schmitt}, {Schubert}, {Sturm}, {Thiel}, {Tofani}, {Vavrek}, {Wetzstein},
  {Wieprecht}, \& {Wiezorrek}}]{pogli10}
{Poglitsch}, A., {Waelkens}, C., {Geis}, N., {et~al.} 2010, \aap, 518, L2

\bibitem[{{Rawlings} {et~al.}(2004){Rawlings}, {Redman}, {Keto}, \&
  {Williams}}]{raw04}
{Rawlings}, J.~M.~C., {Redman}, M.~P., {Keto}, E., \& {Williams}, D.~A. 2004,
  \mnras, 351, 1054

\bibitem[{{Robitaille} {et~al.}(2007){Robitaille}, {Whitney}, {Indebetouw}, \&
  {Wood}}]{robi07}
{Robitaille}, T.~P., {Whitney}, B.~A., {Indebetouw}, R., \& {Wood}, K. 2007,
  \apjs, 169, 328

\bibitem[{{Torrelles} {et~al.}(1992){Torrelles}, {Rodriguez}, {Canto},
  {Anglada}, {Gomez}, {Curiel}, \& {Ho}}]{torrelles92}
{Torrelles}, J.~M., {Rodriguez}, L.~F., {Canto}, J., {et~al.} 1992, \apjl, 396,
  L95

\bibitem[{{Ukidss}(2012)}]{lucas12}
{Ukidss}, C. 2012, VizieR Online Data Catalog, 2316, 0

\bibitem[{{Weigelt} {et~al.}(2006){Weigelt}, {Beuther}, {Hofmann}, {Meyer},
  {Preibisch}, {Schertl}, {Smith}, \& {Young}}]{wei06}
{Weigelt}, G., {Beuther}, H., {Hofmann}, K.-H., {et~al.} 2006, \aap, 447, 655

\bibitem[{{Wienen} {et~al.}(2012){Wienen}, {Wyrowski}, {Schuller}, {Menten},
  {Walmsley}, {Bronfman}, \& {Motte}}]{wienen12}
{Wienen}, M., {Wyrowski}, F., {Schuller}, F., {et~al.} 2012, \aap, 544, A146

\bibitem[{{Wu} {et~al.}(2004){Wu}, {Wei}, {Zhao}, {Shi}, {Yu}, {Qin}, \&
  {Huang}}]{wu04}
{Wu}, Y., {Wei}, Y., {Zhao}, M., {et~al.} 2004, \aap, 426, 503

\end{thebibliography}

\end{document}